\documentstyle[12pt]{article}    
\begin{document}           
\title{Quasi-Two-Dimentional Modeling of the Self Gravitating Gas \thanks{
Cosmoparticle physics. 1. Proceedings of 1 International
conference on cosmoparticle physics "Cosmion-94", dedicated to 80
Anniversary of Ya.B.Zedovich and 5 Memorial of A.D.Sakharov. Eds
M.Yu.Khlopov, M.E.Prokhorov, A.A.Starobinsky, J.Tran Thanh Van, Editions
Frontieres, 1996. p.p.203 - 206.
}}
\author{A. Gromov\thanks{
Computer Science Chair, Faculty of Technical Cybernetics,
St. Petersburg State Technique University,
Dom 29, Politechnitheskaja Ulitsa, 192251,
St. Petersburg, RUSSIA,
E-mail gromov@natus.stud.pu.ru
} ,~V. Perepelovsky \thanks{
Department of Electron-Ion Processing of Solids,
St.Petersburg Electrotechnical University,
Prof. Popov str.,5 St.Petersburg, 197376, Russia,
FAX  +7-(812)-234-9983,
E-mail borisk@sovam.com
}}
\date{}

\maketitle

%
%
%
%
%

\begin{abstract}

The quasi-two-dimensional modeling of the small adiabatic
perturbation on the background of the stationary configuration
of the selfgravitating gas with the weak transverse
nonhomogeneity approximation is presented. The space periodic
character of the solution of this system is proofed.
\end{abstract}

\section{The Physical Model}
%
An effects arising during an expansion of the plane
stationary adiabatic wave with a small amplitude on the background of the
stationary configuration of the self gravitating gas have a volume nonlinear
character.
These effects arising in the process of competition of the
two basic mechanisms:
the self gravitating and thermal  expansion of particles.
Three physical parameters define the solution of this problem:
$\Phi_0$ - the characteristic gravitational potential,
$T_0$ - the characteristic temperature for the thermal expansion and $\gamma$
- the index of adiabatic. The stationary configuration of the
self gravitating gas has cylindrical symmetry \cite{A.D.}.
We use the term "radius" to fix the boundary between the
areas of the gas and vacuum.
The  wave  under consideration perturbed to a small extent the
initial distribution of the speed, density and radius. Smallness of the
perturbation is of principle character.
The mathematical description of this model is represen\-ted by the system
of equation of motion and continuity (three dimensional nonstationary
partial differential equations), Poisson equation and algebraic
equation of state:
\begin{eqnarray}
\frac{\partial \vec \upsilon}{\partial t}+(\vec \upsilon \cdot \nabla)
\vec \upsilon&=&
-\frac{\nabla P}{\rho}-\nabla \Phi; \\
\frac{\partial \rho}{\partial t}+\nabla \cdot (\rho \vec \upsilon)&=&0;
\\
{\nabla}^2 \Phi&=&4\pi G \rho; \\
P&=&A\rho^{\gamma},
\label{INITIAL_Dimention}
\end{eqnarray}
there $P$ is pressure, $\vec \upsilon$ - thermal$ $ speed, $\Phi$  -
gravitational
potential, $\rho$ - gas density, $t$ - time,
$A=const$.
This system describes the ideal classic gas with self gravitation.
In the system of coordinate related to the stationary wave
the derivative of time disappear \cite{A.D.} from the system
(\ref{INITIAL_Dimention}).
An addition to (\ref{INITIAL_Dimention})
we'll use also the radius component of the equation of motion
multiplyed by $\rho(z,r)r$:
\begin{equation}
\frac{1}{2}\rho r \frac{\partial \upsilon_r^2}{\partial r}+
r\frac{\partial P}{\partial r}=-\rho r \frac{\partial \Phi}{\partial r}.
\label{R-motion_Dimention}
\end{equation}
\section{The Relations of the Quasi-Two-Dimen\-tional Averaging Method}
%
The QTDA method is
applied to finite three dimensional space problem with separate direction.
A   rigorous   and   consistent   quasi-two-dimensional
approach must based  on  the cross-section  averaging  of  three dimensional
physical model equations. This procedure  allows  one
to  proceed  from  three-dimensional  distributions  of  physical
quantities (such as potential,  mass density  etc.)  to  their
one-dimensional averaged distributions \cite{V.V}. The following relations
of the QTDA method have been used:
\begin{eqnarray}
\int \limits_{S(z)}{\nabla}^2 \Phi (r,z) ds&=&\frac{d}{dz}\biggl\{
\biggl(\frac{d\Phi(z)}{dz}\biggr)^{S}S(z)      \biggr\} +
\oint \limits_{L(z)}\frac{\vec n \cdot \nabla \Phi(r,z)}{\cos \vartheta}
dl, \\
\label{QTDAP}
\int \limits_{S(z)}\nabla \Phi (r,z) \cdot d\vec s&=&\frac{d}{dz}\biggl\{
\biggl(\Phi(z)\biggr)^{S}S(z)      \biggr\} +
\oint \limits_{L(z)}\frac{\vec n \cdot \vec {e_z} \Phi(r,z)}{\cos \vartheta}
dl, \\
\int \limits_{S(z)}\nabla \cdot \vec A(r,z) ds&=&\frac{d}{dz}\biggl\{
\biggl(A_z (z)\biggr)^{S}S(z)      \biggr\} +
\oint \limits_{L(z)}\frac{\vec n \cdot \vec A(r,z)}{\cos \vartheta} dl,
\end{eqnarray}
where $\vec n$ - external (as referred to the section $S(z)$ ) normal to
the side surface;
$\vartheta$ is the angle between the axis r and the normal $\vec n$ to the
side surface of the cylinder;
\begin{eqnarray}
\biggl(\frac{d\Phi(r,z)}{dz}\biggr)^{S}S(z)&=&
\int \limits_{S(z)}\nabla\Phi(r,z) \cdot d\vec s; \\
\bigl(\Phi(z)\bigr)^{S}S(z)&=&\int \limits_{S(z)} \Phi(r,z) ds,
\label{def1}
\end{eqnarray}
where index ${}^S$ marks the cross section averaging values.
All characteristics of the three dimensional problem are described in the new
equations
by the relation establishment between new functions - the average quantity
variable of physical value and the boundary meanings of the initial function.
In this formulas the boundary meanings are represented by the contour
integrals. It is seen that all non-one dimensional effects are located in the
contour integrals.
\section{The QTD Model of the Weak Transverse
Nonhomogeneity Approximation}
%
The QTDA equations are the result of application of the QTDA method represented
by $(3)$ - (\ref{def1}) to equations of continuity, Poisson,
and $(2)$ in the system of coordinate related to the wave.
In this model we studing the effects related to
the small perturbations of the physical values. According to this the
smallness of following parameters have been supposed:
\begin{eqnarray}
\mu&=&-\frac{\upsilon^2_r(R)-(\upsilon^2_r)^{S}}{c_0^2},
\\
d&=&\frac{(\rho^{\gamma})^{S}-\bigl({\rho^{S}}\bigr)^{\gamma}}{\rho_0^{\gamma}},
\label{small_density}
\end{eqnarray}
where $c_0$ and $\rho_0$ are the sound speed and density on the
axis z in the nonperturbated part of the cylinder.
The smallness of these parameters correspond to the weak transverse
nonhomogeneity approximation.
We study the equations of continuity, Poisson, state and $(2)$
in the $0$-approximation of small parameters (\ref{small_density}).
The structure of the QTDA equations depended from contour integrals,
which are simplified in the $0$-approximation because in this case:
$1)$ $\cos \theta=1$;
$2)$ $\vec n$ is parallel to $\nabla\Phi$; $\vec n\cdot \vec e_z=0$.
Therefore only the Poisson equation has a nontrivial summand after the
averaging.
Thus the QTDA equations under consideration become as follows:
\begin{equation}
\frac{1}{2}\bigl(\upsilon^2\bigr)^{S}=
-\frac{A\gamma}{\gamma-1}\bigl(\rho^{S}\bigr)^{\gamma-1}-\Phi^{S}+C,
\label{11}
\end{equation}
where $C=const$;
equation of continuity: $\rho^{S} \upsilon^{S}=const$;
Poisson equation:
\begin{equation}
\frac{d^2\Phi^{S}(z)}{dz^2}-\frac{2}{R^2}\Phi^{S}(z)=4\pi G \rho^{S}(z).
\end{equation}
It is not possible to obtain the second summand from one dimensional model.
The averaging of the 0-approximation of the equation $(2)$ represent the
dependence of the radius of the cylinder from the coordinate $z$:
\begin{equation}
\frac{dR}{dt}=\upsilon_r(R)=\upsilon_z K(\sigma)\frac{d\sigma}{dz},
\end{equation}
where $\sigma$ is the surface density, $K(\sigma)=
\frac{2-\gamma}{2(1-\gamma)}\sigma^{\frac{2-\gamma}{2(1-\gamma)}-1}$.
Let's estimate benefit of transverse motion energy in the full energy of
the system.
For this purpose we'll substitute
$\vec \upsilon=\vec \upsilon_r+\vec \upsilon_z$
in (\ref{11}) and compare the energy of longitudinal and transversements
we find out, that the latters are unessential far from the wave front
$\triangle z$,
\begin{equation}
\triangle z \gg max \left|{\frac{\upsilon_z(R)K(\sigma)}{\sqrt{F_i}}}
\triangle \sigma\right|,
\label{?/?}
\end{equation}
where $F_i$ - one of the set from
$\biggl({\bigl(\upsilon_z^2\bigr)^{S},
\frac{2A\gamma}{\gamma-1}\bigl(\rho^{S}\bigr)^{\gamma-1},
2\left|{\Phi^{S}}\right|,
2\left|{C}\right|
\biggr)}.$
$  \triangle \sigma$ - is a
changing of surface density along the length $\triangle z$.
It means that volume effects are more valuables than valuable surface ones.
We'll use in following the averaging values without index ${}^S$.
\section{The QTD Description of the Self Gravitating Gas}
%
For future investigation the dimensionless variables
(the coordinate, density, pressure and potential) have been introduce:
\begin{eqnarray}
\xi&=&\frac{r}{L_j}, \ \ \ \ \
\delta=\frac{\rho}{\rho_0}, \ \ \ \ \
p=\frac{P}{P_0}, \ \ \ \ \
\varphi=\frac{\Phi}{\Phi_0}, \\
L_j^2&=&\frac{{\pi}c_0^2}{G\rho_0}, \ \ \ \ \
{c_0}^2=\Biggl(\frac{dp}{d{\rho}}\Biggr)_0=A{\gamma}{\rho_0}^{\gamma-1},
\label{Dimensionlesses}
\end{eqnarray}
where $L_j$ - is the Djeence's length, $c_0$ - sound speed.
There are characteristic values marked with index "0" in this formulas.
The dimensionless equations are:
\begin{eqnarray}
\dot \varphi=u\bigl(1-\gamma \delta^{\gamma+1}\bigr), \\
\dot u=(\varphi+\delta)\bigl(1-\gamma \delta^{\gamma+1}\bigr), \\
\dot \delta=\beta u\delta^3,
\label{Dimentionless_QTDA}
\end{eqnarray}
where   $\dot {}=\frac{d}{d \tau}, \ \ \ \ d\xi=
(1-\gamma) \delta^{\gamma+1} d \tau$.
The meaning of density related to sound speed is
$\delta_s=\biggl({\frac{1}{\gamma}}\biggr)^{\frac{1}{\gamma+1}}$, so the
(\ref{Dimentionless_QTDA}) is not equal to infinity.
The first integral of (\ref{Dimentionless_QTDA}) is
\begin{eqnarray}
\varphi=g(\delta)+C, \\
g(\delta)=-\frac{1}{\beta}\biggl(\frac{1}{2\delta^2}+
\frac{\gamma}{\gamma+1}\delta^{\gamma-1}\biggr),
\label{1_int}
\end{eqnarray}
where $C=const$. From the system of equations
\begin{eqnarray}
\frac{du}{d\tau}=(1-\gamma\delta^{\gamma+1})(g(\delta)+\delta+C), \\
\frac{d\delta}{d\tau}=\beta u \delta^3,
\label{15}
\end{eqnarray}
the second integral of the system (\ref{15}) has been received and
has a form of "energy conservation law":$E=\frac{u^2}{2}+U$,
where $u$ plays a role of speed, $E$ is the constant "energy", and
the "potential"
\begin{equation}
U=-\frac{1}{2}[C+g(\delta)]^{2}+
\frac{1}{\beta}\bigl(\frac{1}{\delta}+\delta^{\gamma}\bigr).
\label{U}
\end{equation}
The local minimum of the potential $U$ correspond to finite solutions of
the system (\ref{15}). It is according to result have received in \cite{A.D.}
for $\gamma=2$.
\section{Acknowledgements}

I'm grateful to Prof. Arthur D. Chernin for encouragement and
discussion. This paper was financially supported by "COSMION" Ltd., Moscow.
\end{document}